\documentclass[a4paper,14pt]{extarticle}
\usepackage{cmap} 
\usepackage[cp1251]{inputenc}
\usepackage[english,russian]{babel}
\usepackage{braket,amssymb,amsmath,array,commath,mathtext,wrapfig,tikz}
\usepackage[arrows=pgf-filled,version=3]{mhchem}
\usepackage{bm}
\usepackage{amsthm,amsfonts,amscd}
\usepackage{graphicx}
\usepackage{cite}
\usepackage{lscape}
\usepackage{geometry}
\usepackage{indentfirst}
\usepackage{afterpage,soul,ulem,dcolumn,changebar,longtable,hhline,multirow,makecell}
\usepackage[small,bf]{caption}
\makeatletter
\renewcommand{\@biblabel}[1]{#1.\hfil} 
\makeatother

\begin{document}
\renewcommand{\refname}{References}
\begin{center}
\textit{Generalized Vaidya spacetime: horizons, conformal symmetries, surface gravity and diagonalization}
\end{center}
	\begin{center}
 \textbf{Vitalii Vertogradov}$^{1}$,
 \textbf{Dmitriy Kudryavcev}$^{2}$\\
	\end{center}

	\begin{center}
		$^{1}$ \quad 
  Physics department, Herzen state Pedagogical University of Russia,
		
		48 Moika Emb., Saint Petersburg 191186, Russia
		
		SPb branch of SAO RAS, 65 Pulkovskoe Rd, Saint Petersburg 196140, Russia
		
		vdvertogradov@gmail.com \\
  $^{2}$ \quad 
  Physics department, Herzen state Pedagogical University of Russia,
		
		48 Moika Emb., Saint Petersburg 191186, Russia
 \\ kudryavtsiev33@gmail.com
	\end{center}
\textbf{Abstract:} In this paper, the different properties of generalized Vaidya spacetime are considered. We define the location of horizons. We show that the apparent horizon can contain the event horizon. The locations of all types of horizons are compared with ones in the usual Vaidya spacetime. We investigate the timelike geodesics in this spacetime. New corrections to Schwarzschild and Vaidya cases appear and we give conditions when these corrections are not negligible. Also, we consider the conformal Killing vector and transform the metric to conformally-static coordinates. We introduce a new constant of motion along null and timelike geodesics, which is generated by a homothetic Killing vector. The conformally-static coordinates allow diagonalizing of the generalized Vaidya spacetime. The surface gravity has been calculated for the dust and stiff fluid cases.


\section{Introduction}

Nowadays, the black hole plays an important role in modern theoretical physics and astrophysics. In the real world, astronomical bodies gain mass when they absorb radiation and they lose mass when they emit radiation, which means that the space-time around them is time-dependent. Vaidya solution~\cite{bib:vaidya} is one of the exact dynamical solutions of the Einstein equations. It can be regarded as a dynamical generalization of the static Schwarzschild solution. The~Vaidya
spacetime is widely used in many astrophysical applications with strong
gravitational fields. In~general relativity, this spacetime assumed
added importance with the completion of the junction conditions at the
surface of the star by Santos~\cite{bib:santos1985non}. The~pressure at
the surface is non-zero, and~the star dissipates energy in the form of
heat flux. This made it possible to study dissipation and physical
features associated with gravitational collapse, as~shown by Herrera et al.~\cite{bib:herrera2006some, bib:herrera2012dynamical,bib:joshivaidya}. The question about the dynamical shadow formation in Vaidya spacetime is discussed in~\cite{bib:germany, bib:japan}. The horizon structure and entropy of this solution are investigated  for an empty background in~\cite{bib:nelvaidya, bib:nelsurface}, for Vaidya surrounded by cosmological fields in~\cite{bib:tur1, bib:tur2, bib:tur3}. Some recent studies of the temperature properties inside the radiating
star have been done in~\cite{bib:reddy2015impact, bib:thirukkanesh2012final, bib:thirukkanesh2013}. The Vaidya spacetime can be extended to include both null dust and null string fluids leading to
the generalized Vaidya spacetime~\cite{bib:vunk}. A detailed investigation of the properties of the generalized Vaidya spacetime can be found in~\cite{bib:husain1996exact, bib:Radiationstring1998, bib:twofluidatm1999}. The generalized Vaidya spacetime has been used to investigate gravitational collapse ~\cite{bib:Maombi1, bib:Maombi2, bib:ver1, bib:ver2, bib:ver3, bib:myrev, bib:r2}. The conformal symmetries and embedding and other properties are discussed in~\cite{bib:maharaj, bib:r4, bib:r1, bib:r3}. 

The generalized Vaidya spacetime has the off-diagonal term which can lead to the negative energy for a particle. The absence of such particles has been proven  in~\cite{bib:verneg}. The forces in Vaidya spacetime are discussed in~\cite{ bib:verforce, bib:verlinear}. Recently, a new generalization of Vaidya spacetime has been found by the gravitational decoupling method~\cite{bib:vernew} which can describe the Vaidya black hole distorted by dark matter.

In the present work, we investigate different horizon locations. Two  approaches exist for locating the horizon of a black hole:
\begin{enumerate}
\item  The causal approach, is familiar, especially in the form of the global event horizon.
\item The quasi-local approach, typically based on marginally outer trapped surfaces\footnote{The event horizon is the global notion and it is in principle impossible to locate it. One of the possible alternatives is  to use the trapped surfaces. this trapped surfaces are not entirely local since they are closed spacelike surfaces, these provide a quasi-local alternative which an observer could in principle locate in order to detect the presence of a black hole~\cite{bib:quasi}.}.
\end{enumerate}
Defined horizons in this way coincide in globally static spacetimes such as the Schwarzschild solution but differ often in dynamical spacetimes. The key difference is that the event horizon is always a null surface. The apparent horizon, on the other hand, might be not only null but also spacelike and timelike. The event horizon candidates for slowly evolving charged Vaidya spacetime have been studied in~\cite{bib:newin}. A natural physical phenomenon to associate with the boundary of a black hole is Hawking radiation~\cite{bib:hawkingradation}. However, it is a hard task to define the surface gravity in dynamical spacetime~\cite{bib:nelsurface}.  Here we consider some of the definitions and calculate the surface gravity in the generalized Vaidya spacetime case.

As light can't escape a black hole, the only possibility to investigate its properties is to study its impact on the surrounding matter and its movement. When one considers the Schwarzschild black hole, then geodesics of motion gives a Newtonian gravity force, centrifugal repulsive force, and new general relativity corrections related to precession. Vaidya spacetime doesn't change these quantities but, instead, adds a new one-induced acceleration~\cite{bib:tur1} which is related to  generalized 'total apparent flux'~\cite{bib:124}. This flux can be negative in generalized Vaidya spacetime - the well-known example is - charged Vaidya solution $=$ generalized Vaidya solution with the equation of the state of the stiff fluid. When flux becomes negative, the null energy condition is violated. In this regard, it is really important to know the event or apparent horizon location to understand if these surfaces cover the region  where the null energy condition is violated. 

In the generalized Vaidya spacetime,  new corrections to Schwarzschild and usual Vaidya cases appear. These corrections can be negligible, but they can be large enough to be compared with initial forces and change their orientation. 

One more important question is the symmetries related to the conformal symmetry of spacetime. For the certain choice of the mass function, the Generalized Vaidya metric possesses the homothetic Killing vector. This extra symmetry can help to define the constant of motion related to the angular momentum and energy. We transform the metric to conformally-static coordinates and show the simple method of one can diagonalize generalized Vaidya spacetime, which can play an important role in the interpretation of physical results obtained in this spacetime.

This paper is organized as follows. In sec. II we define the apparent and event horizons and calculate their approximate location for different mass functions. Also, the null surface with constant area change is discussed. The difference in horizon locations between Vaidya and generalized Vaidya cases is discussed. In sec. III we discuss the timelike geodesics. We find conditions when, new corrections to the Vaidya case, are not negligible. In sec. IV We define the homothetic Killing vector and transform the metric to the conformally-static coordinates. Also, we define the new constants of motions of  a particle which  are the results of the conformal symmetry. In sec. V we discuss the different methods of dynamical surface gravity definition in the generalized Vaidya spacetime. The obtained results are compared to the Vaidya spacetime case. Sec. VI has dealt with the diagonalization of generalized Vaidya spacetime. Sec. VII is the conclusion.

The system of units $c=G=1$and signature $-\,, +\,, +\,, +$ will be used throughout the paper. 

\section{Generalized Vaidya horizons}

The generalized Vaidya spacetime in Eddington-Finkelstein coordinates has the following form~\cite{bib:vunk}:
\begin{equation} \label{eq:metric}
\begin{split}
ds^2=-\left(1-\frac{2M(v,r)}{r}\right) dv^2+2dvdr+r^2d\Omega^2 \,, \\
d\Omega^2=d\theta^2+\sin^2\theta d\varphi^2\,.
\end{split}
\end{equation}
Here $M(v,r)$ is the mass function which depends upon the advanced Eddington time $v$ and space areal coordinate $r$ i.e. $r$ is a coordinate such that the surfaces of spherical isometry have area  $A=4\pi r^2$.

The energy-momentum tensor of \eqref{eq:metric} represents the combination of the null dust (type-I) and the null fluid (or null strings, type-II)\footnote{The classification of the energy-momentum one can find in the textbook~\cite{bib:hok}}:
	\begin{equation} \label{eq:ten}
		\begin{split}
T^{(m)}_{ik}= \mu n_{i}n_{k}\,, \\
T^{(n)}_{ik}=(\rho+P)(a_{i}b_{k}+a_{k}b_{i})+Pg_{ik} \,, \\
\mu=\frac{2 \dot{M}}{ r^2} \,, \\
\rho=\frac{2M'}{r^2} \,, \\
			P=-\frac{M''}{r} \,, \\
a_{i}=\delta^0_i \,, \\
b_{i}=\frac{1}{2} \left (1-\frac{2M}{r} \right )\delta^0_{i}-\delta^1_{i} \,, \\
a_{i}a^{i}=b_{i}b^{i}=0 \,, \\
b_{i}a^{i}=-1 \,.
		\end{split}
	\end{equation}
	here $P$ - pressure, $\rho$ - density, $\mu$ - the energy density of the null dust. And $a,b$ - two null vectors. Here $T^{(m)}_{ik}$ is the energy-momentum tensor of null dust and $T^{(n)}_{ik}$ - null fluid.

The energy-momentum tensor \eqref{eq:ten} should satisfy the weak, null, strong and dominant energy conditions. Of course, all these conditions can be violated under some particular circumstances. For example in the case of Hawking radiation the weak and null energy conditions are violated. The strong energy condition is violated if one considers models with dark energy or consider a regular black hole. The dominant energy condition is violated in the case phantom field. 
Strong and weak energy conditions demand:
\begin{equation}
\begin{split}
\mu \geq 0 \,, \\
\rho \geq 0 \,, \\
P\geq 0 \,.
\end{split}
\end{equation}
If $P<0$ then only strong energy condition is violated but the weak and null ones are satisfied.
The dominant energy condition imposes following conditions on the energy momentum tensor:
\begin{equation}
\begin{split}
\mu \geq 0 \,, \\
\rho \geq P\geq 0\,.
\end{split}
\end{equation}

To calculate a future outer trapping horizon (FOTH), one needs to know the radial outgoing $\Theta_l$ and ingoing $\Theta_n$ null expansions. The apparent horizon exists if at some radius $r_{ah}$ the following conditions are held:
\begin{equation} \label{eq:conah}
\Theta_l(r_{ah})=0 \,, \Theta_n(r_{ah})<0 \,.
\end{equation}

If $V^i$ is the affinely parameterized tangent vector to the geodesic congruence, then the expansion $\Theta_V$ can be defined as $V^i_{;i}$. The expansion for non-affinely parameterized vector in generalized Vaidya spacetime has been calculated in~\cite{bib:ver1}.

In generalized Vaidya spacetime \eqref{eq:metric}, the expansions are given by:
\begin{equation} \label{eq:ah}
\begin{split}
\Theta_l=\frac{1}{r} \left(1-\frac{2M(v,r)}{r}\right) \,, \\
\Theta_n=-\frac{2}{r} \,.
\end{split}
\end{equation}

From \eqref{eq:ah} one can see that if we satisfy the conditions \eqref{eq:conah}, then we obtain the following apparent horizon equation:
\begin{equation} \label{eq:ahrad}
r_{ah}=2M(v,r_{ah}) \,.
\end{equation}

If we consider the equation of the state $P=\alpha \rho$ for the type-II matter field, then, by virtue of the Einstein equations, the mass function is given by~\cite{bib:ver2, bib:ver3, bib:vunk}:
\begin{equation} \label{eq:mass}
\begin{split}
M(v,r)=C(v)+D(v)r^{1-2\alpha} \,, \\
\alpha \in [-1\,, 1] \,, \alpha \neq \frac{1}{2} \,.
\end{split}
\end{equation}
Where $C(v)$ and $D(v)$ are arbitrary functions of the Eddington time $v$. If  $D(v)\equiv 0$ then one has the usual Vaidya solution and \eqref{eq:ahrad} gives the well-known result $r=2M(v)=2C(v)$. In the dust case  $\alpha=0$ \eqref{eq:ahrad} gives~\cite{bib:ver1}:
\begin{equation} \label{eq:dustah}
r_{ah}=\frac{2C(v)}{1-2D(v)} \,.
\end{equation}

In this case the energy conditions~\cite{bib:pois} demands $C(v) \geq 0 \,, D(v)\geq 0$. Hence, one can conclude if $D(v)\neq 0$ then the location of the apparent horizon in this case doesn't coincide with the apparent horizon in Vaidya spacetime. Moreover, $r_{ah}^{GeneralizedVaidya}>r_{ah}^{Vaidya}$. The apparent horizon area in the dust case is given by:
\begin{equation}
A^{dust}=\frac{16\pi C^2(v)}{(1-2D(v))^2} \,.
\end{equation}

Cases when $D(v)\geq \frac{1}{2}$ is considered in~\cite{bib:ver1}.

If the type-II represents the stiff fluid~\cite{bib:zel} $\alpha=1$, then we have two apparent horizons:
\begin{equation} \label{eq:stiffah}
r^{\pm}_{ah}=C(v)\pm\sqrt{C^2(v)+2D(v)} \,.
\end{equation}
Here, the energy conditions demand $C(v)\geq 0\,, D(v)\leq 0$~\cite{bib:etnak}. We are interested in outer horizon $r^+_{ah}$. In comparison with the dust case $r^{generalized\_Vaidya}_{ah}<r^{Vaidya}_{ah}$. As the result, the area of the outer apparent horizon is:
\begin{equation}
A^{stiff\,fluid}=4\pi(C(v)+\sqrt{C^2(v)+2D(v)})^2 \,.
\end{equation}

We can conclude that the location of the apparent horizon depends upon the function $D(v)$. The sign of this function depends upon the energy conditions. The week energy condition demands the positivity of the energy density $\rho$. Substituting \eqref{eq:mass} into \eqref{eq:ten}, one obtains the following energy density expression:
\begin{equation} \label{eq:loc}
\rho=2\left (1-2\alpha \right)\frac{D(v)}{r^{2+2\alpha}}\geq 0 \,.
\end{equation}

From \eqref{eq:loc} one can see that we have two cases:
\begin{equation}
\begin{split}
\alpha<\frac{1}{2} \rightarrow D(v)\geq 0 \longrightarrow r^{generalized\_Vaidya}_{ah}>r^{Vaidya}_{ah} \,, \\
\alpha>\frac{1}{2} \rightarrow D(v)\leq 0 \longrightarrow r^{generalized\_Vaidya}_{ah}<r^{Vaidya}_{ah} \,.
\end{split}
\end{equation}

It is a hard task to find the location of the event horizon in dynamical spacetimes. Here, we will consider to what extent strictly null horizons can be defined quasi-locally. 

The event horizon is defined as connected components of the past causal boundary of future null infinity and is generated by null geodesics that fail to reach infinity. The event horizon is always a null surface since it is a causal boundary. In generalized Vaidya spacetime \eqref{eq:metric} the coordinate $v$ is constant on ingoing radial null geodesics. Any outgoing radial null geodesic must satisfy:
\begin{equation} \label{eq:nulvec}
\frac{dr}{d\lambda}=\frac{1}{2}\left(1-\frac{2M}{r}\right)\frac{dv}{d\lambda} \,,
\end{equation}
for some affine parameter $\lambda$. 

We consider the metric \eqref{eq:metric}for accreting matter. In this case, one can also find the approximate location of the event horizon by imposing the condition:
\begin{equation} \label{eq:event}
\frac{d^2v}{dr^2}=0 \,.
\end{equation}

This formula means that the event horizon is evolving at a steady rate. 
Taking the second derivative \eqref{eq:nulvec}:
\begin{equation} 
\frac{d^2r}{d\lambda^2}=\frac{1}{2}\left(1-\frac{2M}{r}\right)\frac{d^2v}{d\lambda^2}+\left(\frac{M}{r^2}-\frac{M'}{r}\right)\frac{dv}{d\lambda}\frac{dr}{d\lambda}-\frac{\dot{M}}{r}\left(\frac{dv}{d\lambda}\right)^2 \,,
\end{equation}
and imposing  the condition 
\begin{equation} \label{eq:new1}
\frac{d^2v}{d\lambda^2}=0 \,, \frac{d^2r}{d\lambda^2}=0 \,,
\end{equation}

(Which follows from \eqref{eq:event}) one finds:
\begin{equation} \label{eq:long1}
\left(\frac{M}{r^2}-\frac{M'}{r}\right)\frac{dv}{d\lambda}\frac{dr}{d\lambda}-\frac{\dot{M}}{r}\left(\frac{dv}{d\lambda}\right)^2=0 \,.
\end{equation}
And using \eqref{eq:nulvec}, one obtains:
\begin{equation} \label{eq:dop}
\left(\frac{1}{2}\left(1-\frac{2M}{r}\right)\left(\frac{M}{r^2}-\frac{M'}{r}\right)-\frac{\dot{M}}{r}\right)=0 \,.
\end{equation}

And approximate location of the event horizon in the generalized Vaidya spacetime \eqref{eq:metric} is given by:
\begin{equation} \label{eq:eventsol}
(2\dot{M}+M')r^2-M(1+2M')r+2M^2=0 \,.
\end{equation}
The solutions of \eqref{eq:eventsol} allow us to define the location of the event horizon quasi-locally. The equation \eqref{eq:eventsol}, in general, has a lot of solutions but only the outermost horizon is of immediate interest and comments will be restricted mostly to that case. 

We can't find the general solution of \eqref{eq:eventsol} because the mass function also depends upon $r$ coordinate. Here, we again consider two cases of the dust and stiff fluid. Let's suppose again that the mass function is in form \eqref{eq:mass}. we assume that functions $C(v)$ and $D(v)$ in the dust case has the particular form:
\begin{equation} \label{eq:cddust}
\begin{split}
C(v)=\nu v \,, \nu>0 \,,
D(v)=\mu= const. \,, \mu>0 \,.
\end{split}
\end{equation}

In Vaidya spacetime, one has the following restriction $\nu\leq\frac{1}{16}$~\cite{bib:nelvaidya}. Also, substituting \eqref{eq:cddust} into \eqref{eq:dustah}, one obtains the restriction $\mu<\frac{1}{2}$. By using \eqref{eq:cddust} in \eqref{eq:eventsol}, one has the equation for defining the approximate location of the event horizon:
\begin{equation}
2r^2-(1-2\mu)vr+2\nu v^2=0 \,,
\end{equation}
which gives us the radius $r_{eh}$ of the event horizon:
\begin{equation} \label{eq:radehdust}
r_{eh}=\frac{v}{4}\left(1-2\mu+\sqrt{(1-2\mu)^2-16\nu}\right) \,.
\end{equation}
We have dropped the minus sign because the inner event horizon is hidden for distant observer.

To get the approximate location of the event horizon in Vaidya spacetime, one needs to put $\mu=0$~\cite{bib:nelvaidya}. Again, like in the apparent horizon case above, the event horizon in usual Vaidya is bigger than the event horizon radius in generalized Vaidya spacetime in the dust case ($r^{gv}_{eh}<r^v_{eh}$). Also one should note, that the apparent horizon is inside the event horizon ($r_{ah}<r_{eh}$). However, if either $\nu>\frac{1}{16}$ or $(1-2\mu)^2<16\lambda$, then the event horizon is absent. 

Although, we have stated that only outer horizon is of immediate interest, one should realize that if we put $v=const.$ then the radius of the outer horizon tends to infinity and the inner horizon 
\begin{equation}
r^-_{eh}\approx \frac{2\nu v_{c}}{1-2\mu} +O(\nu^2) \,,
\end{equation}
in the limit $\mu\rightarrow 0$ becomes the Schwarzschild event horizon ($v_{c}$ is some positive constant and $M=\nu v_{c}$). 

Figure \ref{fig:1} shows that $r_{eh}\geq r_{ah}$ and $r_{eh}=r_{ah}$ only at $v=0$.
\label{fig:1}
This figure shows the linear behaviour of the event and apparent horizon. Here $\mu=1/64$ and $\nu=1/32$.

\begin{figure}[h]

\centering

\includegraphics[width=0.8\linewidth]{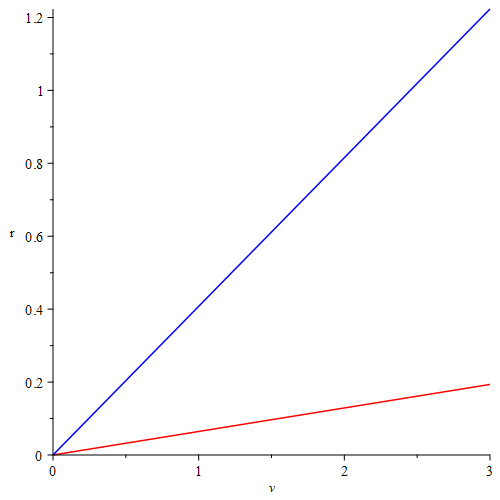}

\caption{This figure is plotted in the assumption $\mu=-1/128$ and $\nu=1/4$. Red line is the apparent horizon and the blue one is the event horizon.}

\label{fig:1}

\end{figure}

If we consider the stiff fluid, i.e. $\alpha=1$ then we can define the functions $C(v)$ and $D(v)$ in the following way:
\begin{equation} \label{eq:cdstiff}
\begin{split}
C(v)=\nu v \,, \nu>0 \,, \\
D(v)=\mu v^2 \,, \mu <0 \,.
\end{split}
\end{equation}

Now if we put \eqref{eq:cdstiff} into \eqref{eq:dop}, then we obtain:
\begin{equation} \label{eq:stifsol}
\left(1-\frac{2\nu v}{r}-\frac{2\mu v^2}{r^2}\right)\left(\frac{\nu v}{r}+\frac{2\mu v^2}{r^2}\right)-2\nu-\frac{4\mu v}{r} = 0 \,.
\end{equation}

One of the solution of \eqref{eq:stifsol} is 
\begin{equation} \label{eq:dusteh}
r_{eh}=-\frac{2\mu v}{\nu} \,.
\end{equation}
The apparent horizon for the choice  of functions $C(v)$ and $D(v)$ \eqref{eq:cdstiff} is
\begin{equation} \label{eq:wdustah}
r_{ah}^{\pm}=\left(\nu\pm \sqrt{\nu^2+2\mu}\right) v\,.
\end{equation}

For the existence both and the apparent horizon \eqref{eq:wdustah} and the event horizon \eqref{eq:dusteh}, the following condition $-\frac{\nu^2}{2}\leq \mu \leq 0$ must be held. If we satisfy this condition then the event horizon is hidden inside the apparent one. If this condition is violated, then one has only the event horizon. 

\begin{figure}[h]

\centering

\includegraphics[width=0.8\linewidth]{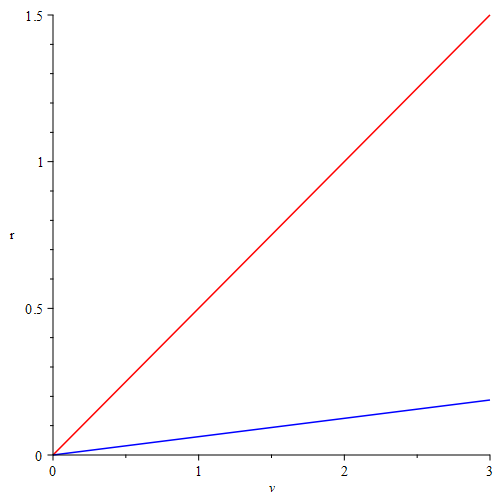}

\caption{This figure is plotted in the assumption $\mu=-1/128$ and $\nu=1/4$. Again, like in the figure \ref{fig:1} the red line shows the behaviour the apparent horizon and the blue line is the event one.}

\label{fig:2}

\end{figure}

Figure \ref{fig:2} demonstrates that in this case the apparent horizon contains the event horizon.

If one considers the case $\nu^2=-2\mu$ then the event horizon location is:
\begin{equation} \label{eq:rn}
r_{eh}= \nu v \,,
\end{equation}

One should realize that the location of the event horizon \eqref{eq:dusteh} is defined quasi-locally. However, if the condition $\nu^2=-2\mu$ is held and we consider the case $v=v_{rn}=const.$, then \eqref{eq:rn} coincides with the event horizon of the extremal Reissner--Nordstr\"om black hole.

We have considered two models. The dust case confirms the state that the apparent horizon is hidden inside the event one~\cite{bib:nelspat}. The stiff fluid model, on the other hand, contradicts to this statement because we have shown that the event horizon is inside the apparent one.

To conclude this chapter, let's find the approximate location of the null surfaces which parameter rate of the area change is constant. The change in the surface area $A=4\pi r^2$ along outgoing null geodesic congruence is given by:
\begin{equation} \label{eq:areadef}
    \frac{dA}{d\lambda}=4\pi r\left(1-\frac{2M}{r}\right)\frac{dv}{d\lambda} 
\end{equation}

We remind that for any parameterization $\lambda$, the following condition  $\frac{dr}{d\lambda}=\frac{1}{2}\left(1-\frac{2M}{r} \right) \frac{dv}{d\lambda}$ is always held. 

Now, we consider the second derivative  of \eqref{eq:areadef}:
\begin{equation}
 \frac{d^2A}{d\lambda^2}=4\pi \left[\left(\frac{1}{2}\left(1-\frac{2M}{r}\right)(1-2M')-2\dot{M}\right)\left(\frac{dv}{d\lambda}\right)^2+r\left(1-\frac{2M}{r}\right)\frac{d^2v}{d\lambda^2}\right]   
\end{equation}

the location of null surfaces with constant area change $\frac{d^2A}{d\lambda^2}=\frac{d^2v}{d\lambda^2}=0$ is defined by the equation:
\begin{equation} \label{eq:constantsol}
(r-2M)(1-2M')-4\dot{M}r=0 \,.
\end{equation}

Here, we consider only the dust case. For the dust case, we use the mass function definition \eqref{eq:cddust}. In this case the equation \eqref{eq:constantsol} gives:
\begin{equation} \label{eq:dustcons}
r_{ac}=\frac{2\nu v(1-2\mu)}{1-2\mu-4\nu} \,.
\end{equation}
From this equation we obtain a new restriction $1-2\mu-4\nu>0$. It is easy to show that if we can define the outer event horizon, then the null surface of constant area change is hidden inside it. However, if we compare the location of the apparent horizon $r_{ah}$ and the constant area change surface $r_{ac}$ then they can quasi-locally coincide  or contain each other. All three options depend upon parameters $\nu$ and $\mu$.
\begin{enumerate}
\item Two horizons quasi-locally coincide. In this case  $\mu =\mu^{\pm}$. Where
\begin{equation}
\mu^{\pm}=\frac{1}{4}\left ( 1 \pm \sqrt{1-16\nu} \right) \,.
\end{equation}
\item The apparent horizon is hidden inside of the constant area change null surface. In this case $0\leq \mu <\mu^-<\frac{1}{2}$ and $\mu^+< \mu <\frac{1}{2}$. 
\item The null surface with constant area change is inside the apparent horizon. In this case $0\leq \mu^- < \mu < \mu^+ < \frac{1}{2}$.
\end{enumerate}

\section{Timelike geodesics}

We can restrict our consideration of the timelike geodesics by the equatorial plane ($\theta=\frac{\pi}{2}$) because of the spherical symmetry all geodesics lie in the plane. There are several ways how to obtain the second order geodesic equation, but we will consider and   vary the following action:
\begin{equation} \label{eq:action}
\mathcal{S}=\int \mathcal{L}d\tau=\frac{1}{2}\int \left[-\left(1-\frac{2M}{r}\right)\left(\frac{dv}{d\tau}\right)^2+2\frac{dv}{d\tau}\frac{dr}{d\tau}+r^2\left(\frac{d\varphi}{d\tau}\right)^2 \right] d\tau \,.
\end{equation}

Due to spherical symmetry, the generalized Vaidya spacetime admits spacelike Killing vector $\frac{\partial}{\partial \varphi}$, which leads to the  constant of motion, i.e. the angular momentum per mass$L$:
\begin{equation} \label{eq:geod1}
L\equiv \frac{\mathcal{L}}{d\left (\frac{d\varphi}{d\tau} \right)}=r^2\frac{d\varphi}{d\tau} \,.
\end{equation}

By varying the action \eqref{eq:action}, one obtains two more equations:
\begin{equation} \label{eq:geod2}
\frac{M'r-M}{r^2}\left (\frac{dv}{d\tau}\right)^2 +\frac{L^2}{r^3}-\frac{d^2v}{d\tau^2}=0 \,,
\end{equation}
\begin{equation}\label{eq:geod3}
\frac{d^2r}{d\tau^2}=-\frac{\dot{M}}{r}\left(\frac{dv}{d\tau}\right)^2 -\left(1-\frac{2M}{r}\right)\frac{d^2v}{d\tau^2}+2\frac{M-M'r}{r^2}\frac{dv}{d\tau}\frac{dr}{d\tau}\,.
\end{equation}

By using the condition $g_{ik}u^iu^k=-1$ and  \eqref{eq:geod2} in \eqref{eq:geod3}, one obtains:
\begin{equation} \label{eq:geod4}
\frac{d^2r}{d\tau^2}=-\frac{\dot{M}}{r}\left(\frac{dv}{d\tau}\right)^2 -\frac{M-M'r}{r^2}-\frac{3M-M'r}{r^4}L^2+\frac{L^2}{r^3} \,.
\end{equation}

The first term represents the generalized 'total apparent flux', the second can be associated with Newtonian gravitational force, the third term represents the general relativity corrections, which accounts for the perihelion precession and the fourth, unperturbed, term corresponds to a repulsive centrifugal force. The last term is the same like in Schwarzschild case, and it is value doesn't depend upon the mass of the central objects. Hence, it can change its orientation. The first three terms, on the other hand, can change its orientation and we will look at them deeper.

\subsection{Generalized total apparent flux}

The first term in \eqref{eq:geod4} is absent in Schwarzschild case and appear in Vaidya spacetime. This term corresponds to non-Newtonian gravitational force associated with the dynamics of a black hole.  Follow Y. Heydarzade~\cite{bib:tur1} we call this term as an induced acceleration  $a_i$:
\begin{equation}\label{eq:ind}
a_i=-\frac{\dot{M}}{r}\left(\frac{dv}{d\tau}\right)^2  \,.
\end{equation}

We can define the generalized total apparent flux ~\cite{bib:124} $\Lambda$, associated with black hole accretion rate, as
\begin{equation}
\Lambda=\dot{M}\left (\frac{dv}{d\tau}\right)^2 \,,
\end{equation}
then, the \eqref{eq:ind} becomes
\begin{equation}
a_i=-\frac{\Lambda}{r} \,.
\end{equation}

Comparison to the Vaidya case shows that $Lambda=0$ not only constant mass $M$ but there is the region where this flux becomes negative. The null energy conditions demand:
\begin{equation}
\Lambda \geq 0 \,.
\end{equation}
Hence, the radius $r_c$, which gives the solution to the equation  $\dot{M}(r_c, v)=0$, is the possible boundary where the null energy condition might be violated. 

If we consider the mass function  \eqref{eq:mass} then one finds
\begin{equation}
\Lambda =\dot{C}+\dot{D}r^{1-2\alpha} \geq 0 \,.
\end{equation}
For $\alpha < \frac{1}{2}$ energy conditions demands $\dot{C}$ and $\dot{D}$ are both positive, hense $\Lambda >0$, however, if $\alpha>\frac{1}{2}$ one must demand $\dot{C}\geq 0$ and $\dot{D}\leq 0$ which leads to the region
\begin{equation}
r< \left(-\frac{\dot{C}}{\dot{D}}\right)^{\frac{1}{1-2\alpha}} \,,
\end{equation}
Where the null energy condition is violated.  This result coincides with~\cite{bib:charged1} for the charged Vaidya spacetime which is the generalized Vaidya spacetime with the stiff fluid $\alpha=1$ and $C(v)=M(v)$ and $D(v)=\frac{Q^2(v)}{2}$ and one obtains the region where the energy conditions are broken:
\begin{equation}
r<\frac{Q\dot{Q}}{\dot{M}} \,.
\end{equation}

However, a particle can't cross this region due to  the Lorentz force~\cite{bib:charged2}.

\subsection{Newtonian gravitational force}

The second term of \eqref{eq:geod4} represents a Newtonian gravitational force. In comparison with Schwarzschild and Vaidya cases, there is a new term $-M'/r$ appears. One should note that if we pick up the mass function in the form:
\begin{equation}
M(v,r)=D(v)r \,.
\end{equation}
Then the term, corresponding Newtonian gravitational force, disappears. The comparison with \eqref{eq:mass} shows that this mass function corresponds to the dust solution with an extra condition $C(v)\equiv 0$. 

In terms of the mass function \eqref{eq:mass}, the Newtonian acceleration $a_n$ takes the form:
\begin{equation} \label{eq:newtonian}
a_n=-\frac{1}{r^2} \left [C(v)+2\alpha D(v)r^{1-2\alpha} \right ] \,.
\end{equation}

Here, one should consider two cases:
\begin{enumerate}
\item $D(v)\geq 0 \rightarrow \alpha <\frac{1}{2}$. If the function $D(v)$ is positive, which energy condition demands, then the only possibility to have the radius at which the \eqref{eq:newtonian} vanishes is negative pressure $\alpha <0$ which leads to the violation of the strong energy condition. In this case the \eqref{eq:newtonian} changes its orientation in the region:
\begin{equation}
r>\left(-\frac{C(v)}{2\alpha D(v)}\right)^{\frac{1}{1-2\alpha}} \,, -1 \leq \alpha < 0 \,.
\end{equation}
\item $D(v)<0 \rightarrow \alpha>\frac{1}{2}$. If one considers $\alpha >\frac{1}{2}$, in particular charged Vaidya spacetime $\alpha=1$, then one has the radius of the sign change of \eqref{eq:newtonian} in the region:
\begin{equation} \label{eq:comp2}
r<\left (-\frac{C(v)}{2\alpha D(v)}\right)^{\frac{1}{1-2\alpha}} \,, \frac{1}{2}< \alpha \leq 1 \,.
\end{equation}

One should note, that orientation change might happen in the region where energy condition is valid and outside the event horizon.
\end{enumerate}

So, for $0\leq \alpha <\frac{1}{2}$, the Newtonian gravitational force never changes its orientation. 

Near black hole, the Newtonian gravitational force can change its orientation  see for example~\cite{bib:force}.

\subsection{General relativity corrections}

The third terms appear in Schwarzschild solution and corresponds to the perihelion precession. Vaidya doesn't change this term, but Generalized Vaidya adds the $-M'/r$ term. 

One should immediately notice that if we pick up the mass function in the form:
\begin{equation}
M(v,r)=D(v)r^3 \,,
\end{equation}
Then the  term, corresponding to general relativity correction
\begin{equation} \label{eq:grcor}
a_{gr}=-\frac{1}{r^4}\left (3M-M'r\right)L^2 \,,
\end{equation}
disappears. For the mass function \eqref{eq:mass} it means that $\alpha=-1$ and $C(v)\equiv 0$. 

For the mass function \eqref{eq:mass}, \eqref{eq:grcor} takes the form:
\begin{equation} \label{eq:grcormass}
a_{gr}=-\frac{L^2}{r^4} \left [3C(v)+(2+2\alpha) D(v) r^{1-2\alpha} \right] \,.
\end{equation}

Like in the previous case, one has two options:
\begin{enumerate}
\item $D(v)>0 \rightarrow \alpha <\frac{1}{2}$. For non-phantom fields, the \eqref{eq:grcormass}disappears at $C(v)\equiv 0$ and $\alpha=-1$. this term can change its orientation only for phantom fields, i.e. $\alpha <-1$. In this case, the region where one might have negative precision is 
\begin{equation}
r>\left ( -\frac{3C(v)}{(2\alpha +2)D(v)}\right)^{\frac{1}{1-2\alpha}} \,.
\end{equation}
\item $D(v)<0 \rightarrow \alpha >\frac{1}{2}$. In this case we have the region with negative precession
\begin{equation} \label{eq:comp1}
r<\left(-\frac{3C(v)}{(2\alpha+2)D(v)}\right )^{\frac{1}{1-2\alpha}} \,.
\end{equation}
\end{enumerate}

The comparison of \eqref{eq:comp1} and \eqref{eq:comp2} shows that the negative precession region contains the Newtonian change orientation region, if the following condition is held:
\begin{equation}
4\alpha -2 > 0 \,,
\end{equation}
which is always valid for our case $\alpha > \frac{1}{2}$. 

\section{Conformal symmetry}

In this section we consider the existence only homothetic Killing vectors which allows us to define an extra constant of motion.

Generalized Vaidya spacetime, in the general case, admits only one constant of motion associated with spherical-symmetry, i.e. angular momentum $L$, which is given by:
\begin{equation}
L=r^2\sin^2\theta \frac{d\varphi}{d\lambda} \,,
\end{equation}
and there are not any other symmetries (in general case) to reduce geodesic equations to the first order differential ones.

However, we can seek for an additional symmetries related to homotetic Killing vector. If a spacetime admits conformal symmetry, then there exists a conformal Killing vector field in the spacetime. If the metric is Lie dragged along this vector field the causal structure of the spacetime remains invariant. 

Any spacetime is said to possess a conformal Killing vector $K^i$ if it solves the following conformal Killing equation:
\begin{equation} \label{eq:confeq}
K_{i;k}+K_{k;i}=c(x^l)g_{ik} \,.
\end{equation}
Where $c(x^l)$ some function which, in general, depends upon all coordinates. If  $c(x^l)=const.$ then $K^i$ is Homothetic Killing vector  and if $c(x^l)\equiv 0$ then $K^i$ is just a Killing vector. The general conformal Killing vector has been considered in~\cite{bib:maharaj}. Usual Vaidya spacetime possesses only Homothetic Killing vector~\cite{bib:frommahhomvay}. Here, we restrict the consideration by Homothetic Killing vector in order to obtain the metrec \eqref{eq:metric} in conformally static coordinates in which the Homothetic Killing vector becomes $\frac{d}{dt}$ which allows us to define a new constant of motion along null geodesics related to the particle energy. 

Let's consider the following vector field:
\begin{equation} \label{eq:vector}
\mathbf{K}=v\frac{d}{dv}+r\frac{d}{dr} \,.
\end{equation}
The vector $K^i$ \eqref{eq:vector} is the Homothetic Killing vector if it satisfies the equation \eqref{eq:confeq} with $c(x^l)\equiv c=const.$ Substituting \eqref{eq:vector} into \eqref{eq:confeq}, we obtain that $K^i$ is the Homothetic Killing vector if the following differential equation is held:
\begin{equation}
    \frac{\dot{M}v}{r}-\frac{M}{r}+M'=0
\end{equation}
The solution of this equation is:
\begin{equation} \label{eq:massconf}
    M(r,v)=\mu r^ \xi v^{1-\xi}+\nu v \,.
\end{equation}
Here $\mu\,, \nu$ and $\xi$ are some constants.  The solution \eqref{eq:massconf} is the mass function \eqref{eq:mass} if:
\begin{equation}
\begin{split}
C(v)=\nu v \,, \nu>0 \,, \\
D(v)=\mu v^{2\alpha} \,,\\
\xi=1-2\alpha \,.
\end{split}
\end{equation}
Where the sign of $\mu$ depends upon the energy condition and $+$ if $\alpha < \frac{1}{2}$ and $-$ if $\alpha > \frac{1}{2}$ and $\alpha$ is the constant from the equation of the state $P=\alpha \rho$. 

Let's define the location of the conformal Killing horizon. This horizon is located at $r=r_{ckh}$ and indicates that the vector $K^i$ becomes null:
\begin{equation}
    k^i k_i=-\left(1-\frac{2M}{r}\right)v^2+2rv=0
\end{equation}

Thus, to obtain the location of the conformal Killing horizon, one needs to solve the following equation:
\begin{equation} \label{eq:radcon}
    2r^2-vr+2Mv=0
\end{equation}

Substituting \eqref{eq:massconf} into \eqref{eq:radcon}, one obtains:
\begin{equation} \label{eq:add}
    2r^2-rv+2\mu r^\xi v^{2-\xi}+2 \nu v^2=0
\end{equation}

In the dust case ($\alpha=0 \rightarrow \xi=1$), we get:
\begin{equation} \label{eq:radckhdust}
r_{ckh}=\frac{v}{4}\left(1-2\mu+\sqrt{(1-2\mu)^2-16\nu}\right) \,.
\end{equation}
 One should note that the conformal Killing horizon location \eqref{eq:radckhdust} coincides with the event horizon location \eqref{eq:radehdust}.  However, by comparing the event horizon \eqref{eq:eventsol} and conformal Killing horizon \eqref{eq:add} equations in general case with the mass function in the form \eqref{eq:massconf}, one can easily see that for different values of $\alpha$ locations of these horizons don't coincide.

Now, to obtain the generalized Vaidya spacetime in conformally-static coordinates, one should do the following coordinate transformation~\cite{bib:germany}:
\begin{equation}
\begin{split}
v=r_{0} e^{\frac{t}{r_0}} \,, \\
r=R e^{\frac{t}{r_0}} \,,
\end{split}
\end{equation}
and substituting it into \eqref{eq:metric}, we have:
\begin{equation} \label{eq:metricconf}
ds^2=e^{\frac{2t}{r_0}}\left [ -\left(1-\frac{2\nu r_0}{R}+\frac{\mu r_0^{2\alpha}}{R^{2\alpha}}-2\frac{R}{r_0}\right)dt^2+2dtdR+R^2d\Omega^2\right] \,.
\end{equation}

In these coordinates the homothetic Killing vector $K^i$\eqref{eq:vector}, becomes:
\begin{equation} \label{eq:convectime}
v\frac{\partial}{\partial v}+r\frac{\partial}{\partial r}=\frac{\partial}{\partial t} \,.
\end{equation}

This homotetic Killing vector is timelike if
\begin{equation} \label{eq:ckh}
1-\frac{2\nu r_0}{R}+\frac{\mu r_0^{2\alpha}}{R^{2\alpha}}-2\frac{R}{r_0}> 0 \,,
\end{equation}
and the location of the  conformal Killing horizon is:
\begin{equation} \label{eq:ckhnc}
1-\frac{2\nu r_0}{R_{ckh}}+\frac{\mu r_0^{2\alpha}}{R^{2\alpha}_{ckh}}-2\frac{R_{ckh}}{r_0}=0 \,.
\end{equation}
From the fact that $\frac{\partial}{\partial t}$ is the homothetic Killing vector, one obtains that the energy $E$:
\begin{equation} \label{eq:conenergynull}
E=e^{\frac{2t}{r_0}}\left[ \left( 1-\frac{2\nu r_0}{R}+\frac{\mu r_0^{2\alpha}}{R^{2\alpha}}-2\frac{R}{r_0} \right ) \frac{dt}{d\lambda}-\frac{dr}{d\lambda}\right] \,, 
\end{equation}

is the conserved quantity along a null geodesic.  The angular momentum now has the following form:
\begin{equation}
L=e^{\frac{t}{r_0}}R^2\sin^2\theta \frac{d\varphi}{d\lambda} \,.
\end{equation}

From the fact that $K^i$ is homothetic Killing vector, one has an additional constant of motion $\varepsilon$ along any type of geodesics. Let $u^i$ is four velocities along the geodesic then:
\begin{equation} \label{eq:homcon}
\varepsilon=K_iu^i-\lambda cg_{ik}u^iu^k \,.
\end{equation}
This constant of motion depends upon the affine parameter $\lambda$. $c$ is the conformal factor \eqref{eq:confeq}. By using \eqref{eq:convectime}, \eqref{eq:conenergynull} and the fact that $c=1$, one has for timelike geodesics
\begin{equation}
-\varepsilon=E-\lambda \,.
\end{equation}

So, conformal symmetry of the generalized Vaidya spacetime allows us to get a new constant of motion related to the fact that the homothetic Killing vector is timelike at the region where the condition \eqref{eq:ckh} is held.

\section{The surface gravity}

In black hole thermodynamics the surface gravity of a black hole plays a role analogous to temperature. However, In a fully dynamical situation, the surface gravity will probably not be directly analogous to a temperature of any thermal spectrum. The surface gravity is likely to play a key role in the emission of Hawking radiation, even in non-equilibrium processes. The surface gravity is usually defined on the Killing horizon. It works well in stationary case, but it breaks down in dynamical situation where there is no the Killing horizon. The key question where one should define the surface gravity in the case, black hole either emits Hawking radiation or accreting matter. Here, we give some well-know definitions and compare them with usual Vaidya spacetime.

The first definition which we will cover is the one  given by Fodor et al~\cite{bib:13}. Let's consider an affinely-parameterized ingoing null geodesic $n^i$ which asymptotic behaviour such that $t_in^i=-1$ where $t^i=(1\,, 0\,, 0\,, 0)$ is an asymptotic Killing vector. This definition works only if the spacetime admits an asymptotically flat spatial infinity. Then for outgoing geodesic $l^i$ one demands that the condition $l^in_i=-1$ is held everywhere in the spacetime. Then the surface gravity is given by:
\begin{equation}
\varkappa_F=-n^il^kl_{k;i} \,.
\end{equation}
For $n^i$ and $L^i$ which we have used to calculate the expansion $\Theta$, one has:
\begin{equation} \label{eq:fodor}
\varkappa_F=\frac{1}{4M} \left (1-2M' \right) \,.
\end{equation}
From energy condition, we have $M'>0$ which leads to the fact that the surface gravity in generalized Vaidya is less than the one in usual Vaidya solution by this definition. Also, if we consider the static limit $M=M(r)$ then the Hawking temperature associated with this surface gravity is less than in Schwarzschild case.

Another definition is associated with Kodama vector $K^i$~\cite{bib:kodama} and has been proposed by Hayward~\cite{bib:14}. The Kodama vector has the property that the combination $K_iT^{ik}$ is divergence free in spherical symmetry. At spatial infinity, it reduces $K^iK_i=-1$. The surface gravity for the apparent horizon is defined by:
\begin{equation}
\frac{1}{2}g^{ik}K^j\left(K_{i;j}-K_{j;i}\right)=\varkappa_KK^k \,.
\end{equation}
In generalized Vaidya solution, the Kodama vector has the form $K^i=(1\,, 0\,, 0\,, 0)$ and we have:
\begin{equation}
\varkappa_K=\frac{1}{4M}\left(1-2M'\right ) \,.
\end{equation}
This surface gravity, in generalized Vaidya solution, coincides with the previous definition given by Fodor et all \eqref{eq:fodor}.

In spherically-symmetric spacetime, one can use the Misner-Sharp mass to define the surface gravity~\cite{bib:11}. We know, that the apparent horizon in generalized Vaidya spacetime is given by:
\begin{equation}
r_{ah}=2M(v,r_{ah}) \,.
\end{equation}
By Differentiating this equation with respect to any parameter $\lambda$ labeling spherically symmetric foliations of the horizon, gives:
\begin{equation}
\frac{dr_{ah}}{d\lambda}=\frac{2M}{dv}\frac{dv}{d\lambda}+\frac{dM}{dr_{ah}}\frac{dr_{ah}}{d\lambda} \,.
\end{equation}
We take $\lambda =v$ and use the fact that $A=4\pi r_{ah}^2$, we obtain:
\begin{equation}
\frac{dM}{dv}=\frac{1}{8\pi} \left(1-2M'\right)\frac{1}{2r_{ah}} \frac{dA}{dv} \,.
\end{equation}
By applying the first law of black hole dynamics $dM=\frac{\varkappa}{8\pi}dA$, one gets:
\begin{equation}
\varkappa=\frac{1}{4M}\left (1-2M'\right) \,.
\end{equation}
Which again coincides with \eqref{eq:fodor}.

One should realize that we are interested only in the case $2M'<1$. If we have $2M'=1$ then the surface gravity vanishes and in this case $r_{ah}$ is so-called putative horizon~\cite{bib:wisser}.

Let's calculate the surface gravity for the cases, dust and stiff fluid. In the dust case, by applying the mass definition \eqref{eq:massconf}, one obtains:
\begin{equation}
\begin{split}
M(v,r_{ah})=\nu v+\mu r_{ah} \,, \\
r_{ah}=\frac{2\nu v}{1-2\mu} \,, \\
\varkappa=\frac{1-2\mu}{4\nu v} \,.
\end{split}
\end{equation}

In the stiff fluid case:
\begin{equation}
\begin{split}
M(v,r_{ah})=\nu v+\mu v^2 r^{-1} \,, \mu <0\,, \\
r_{ah}=\left( nu+\sqrt{\nu^2+2\mu}\right) v \,, \\
\varkappa=\frac{\sqrt{\nu^2+2\mu}}{2\nu v+2\nu v\sqrt{\nu^2+2\mu}+2\mu v} \,.
\end{split}
\end{equation}

\section{The diagonalization of the generalized Vaidya spacetime}

As we have pointed out in the introduction, the Vaidya spacetime \eqref{eq:metric} has a large number of astrophysical and theoretical applications. However, the interpretation of physical results obtained in this metric is complicated because this metric is written in off-diagonal coordinates. The problem is that the null coordinate $v$ is not directly measurable physical quantity. Transition to more physical diagonal coordinates involves analytic difficulties, and the explicit form of the corresponding coordinate transformation is generally unknown~\cite{bib:31ber}. The diagonal form of Vaidya spacetime for linear muss function has been obtained in~\cite{bib:berezin}. Later, the diagonal form for some simple models has been obtained for the generalized Vaidya spacetime~\cite{bib:verdiag}. Here we show the most general coordinate transformation  to the diagonal form in the case if the metric \eqref{eq:metric} possesses the homothetic Killing vector.

Let's consider the metric  in the conformally-static coordinates \eqref{eq:metricconf}. Now we introduce new coordinate $T$  by the relation:
\begin{equation} \label{eq:transform}
t=f(T,R) \,.
\end{equation}
Substituting \eqref{eq:transform} into \eqref{eq:metricconf} one obtains:
\begin{equation} \label{eq:between}
\begin{split}
ds^2=e^{\frac{f(R,T)}{r_0}} \left[ -\delta \left ( \frac{\partial f(R,T)}{\partial T} \right)^2 dT^2+2\frac{\partial f(R,T)}{\partial T} \left ( 1-\delta  \frac{\partial f(R,T)}{\partial R} \right ) dTdR + \right.\\ 
\left.+ \frac{\partial f(R,T)}{\partial R}\left (2-\delta \frac{\partial f(R,T)}{\partial R} \right) dR^2+R^2 d\Omega^2 \right ] \,.
\end{split}
\end{equation}
Here
\begin{equation}
\delta= 1-\frac{2\nu r_0}{R}+\frac{\mu r_0^{2\alpha}}{R^{2\alpha}}-2\frac{R}{r_0} \,.
\end{equation}

We demand that $\frac{\partial f}{\partial T } \neq 0$. From \eqref{eq:between} one can easily see that this metric is in the diagonal form if  $f(R,T)$ satisfies the following differential equation:
\begin{equation} \label{eq:dif}
1-f'\delta =0 \,.
\end{equation}
Further, in the paper dash and dot mean the particular derivative with respect to $R$ and $T$ respectively.

From \eqref{eq:dif} one obtains the solution  in the form:
\begin{equation}
f(R,T)=\int \frac{dR}{\delta}+ E(T) \,.
\end{equation}

Where $E(T)=T$  to ensure that $\dot{f}\neq 0$. Substituting this $f$ into \eqref{eq:between} one obtains the generalized Vaidya spacetime in the diagonal form:
\begin{equation} \label{eq:metricdiag}
ds^2=e^{\frac{2T+2\int \delta ^{-1} dR}{r_0}} \left [-\delta dT^2+\delta^{-1}dR^2+R^2d\Omega^2 \right] \,.
\end{equation}

One should note that this spacetime describes a black hole  and if we put $0\neq E(T)\neq 1$ then $\dot{E}^2$ might best be interpreted as a sort of anomalous redshift that describes how far the total gravitational redshift deviates from that implied by the shape function $M_{s}$~\cite{bib:wisser}:
\begin{equation}
M_{s}(T,R)= e^{\frac{t+\int \delta^{-1}dR }{r_0}}\left( \nu r_0+\mu r_0^{2\alpha}R^{1-2\alpha}\right) \,.
\end{equation}
 
\section{Conclusion}

In this work, we have considered the horizon structure of the generalized Vaidya spacetime.  If the type-II matter field satisfies the equation of the state $P=\alpha \rho$ then the presence of this type of matter field affects the Vaidya solution in the following way:
\begin{itemize}
\item The location of the generalized Vaidya apparent horizon is less than the location of the apparent horizon of the usual Vaidya if $\alpha < \frac{1}{2}$.
\item and vice versa if $\alpha > \frac{1}{2}$.
\end{itemize}

In general, the apparent horizon is hidden inside the event horizon. However, we have shown that there are some models when the event horizon is inside the apparent horizon or it might be absent but the apparent horizon can still exist. 

The comparison of the timelike geodesics in Vaidya and generalized Vaidya revealed the fact that:
\begin{itemize}
\item In generalized Vaidya spacetime, the total apparent flux might be negative which leads to the violation of the weak and null energy conditions.
\item The Newtonian force \eqref{eq:newtonian} might vanish and change its orientation;
\item The Newtonian gravitational force never changes its orientation when $0\leq \alpha <\frac{1}{2}$ (dust and radiation are included in this case);
\item For $\alpha >\frac{1}{2}$ the generalized Vaidya spacetime contains general relativity corrections which lead to the negative precession and this region always contains the region in which the \eqref{eq:newtonian} changes its orientation.
\end{itemize}

The last statement might be very important from the astrophysical point of view because as usual Vaidya and Schwarzschild spacetimes, the negative precession is absent. See for example~\cite{bib:parth} The investigations of such orbits in the case of charged Vaidya spacetime which admits homothetic Killing vector is the question of future research. 

The conformal symmetries of generalized spacetime have been considered in~\cite{bib:maharaj}, here we use the existence of the homothetic Killing vector to transform the generalized Vaidya spacetime to the conformally-static coordinates. Also, these coordinates allow to transform the metric to the diagonal form. This diagonalization includes the obtained results in~\cite{bib:berezin, bib:verdiag}.  We have calculated the location of the conformal Killing horizon and show that it coincides with the event horizon in the dust case but can differ in general. 

There are several ways how one can calculate Hawking temperature. We have considered several models and shown that the Hawking temperature is the largest when only type-I of matter field is presence i.e. in the usual Vaidya case. If one adds null strings, then the temperature is decreasing.

\textbf{acknowledgments}: The author says thanks to grant NUM. 22-22-00112 RSF for financial support. The work was performed within the SAO RAS state assignment in the part "Conducting Fundamental Science Research".

\textbf{Conflict of Interest}: The authors declare that they have no
conflicts of interest.


\begin{thebibliography}{150}
\bibitem{bib:vaidya} Vaidya, P.C. Nonstatic solutions of Einstein's
field equations for spheres of fluids radiating energy. \emph{Phys. Rev.}  \textbf{1951}, \emph{83}, 10.
\bibitem{bib:santos1985non}    Santos, N.O. Non-adiabatic radiating collapse. \emph{Mon. Not. R. Astron. Soc.}  \textbf{1985}, \emph{216}, 403.
\bibitem{bib:herrera2006some}    Herrera, L.; Prisco, A.D.; Ospino, J.Some analytical models of radiating collapsing spheres. \emph{Phys. Rev. D} \textbf{2006}, \emph{74}, 044001.
\bibitem{bib:herrera2012dynamical}     Herrera, L.; Denmat, G.L.; Santos, N.O. Dynamical instability and the expansion-free condition. \emph{Gen. Relativ. Gravit.} \textbf{2012}, \mbox{\emph{44}, 1143.}
\bibitem{bib:joshivaidya} Dwivedi,  I.H.; Joshi, P.S. On the nature of
naked singularities in Vaidya spacetimes Class. \emph{Quantum Grav.} \textbf{1989,} \emph{6}, 1599.
\bibitem{bib:germany} Jay Solanki, Volker Perlick Phys. Rev. D, 105:064056, (2022), arXiv:2201.03274 (gr-qc)
\bibitem{bib:japan} Yasutaka Koga, Nobuyuki Asaka, Masashi Kimura, Kazumasa Okabayashi, \textit{Dynamical photon sphere and time evolving shadow around black holes with temporal accretion}, Phys. Rev. D 105, 104040 (2022) 
\bibitem{bib:nelvaidya} A. B. Nielsen. Revisiting Vaidya horizons. Galaxies, 2:62, 2014.
\bibitem{bib:nelsurface} Nielsen, A.B.; Yoon, J.H. Dynamical surface gravity. Class. Quant. Gravity 2008, 25, 085010, doi:10.1088/0264-9381/25/8/085010.
\bibitem{bib:tur1} Y. Heydarzade, F. Darabi \textit{Surrounded Vaidya black holes: apparent horizon properties}, Eur. Phys. J. C (2018) 78:342
\bibitem{bib:tur2}    Heydarzade, Y.; Darabi, F. Surrounded Vaidya solution by cosmological fields. \emph{Eur. Phys. J. C} \textbf{2018}, \emph{78}, 582.
\bibitem{bib:tur3}    Heydarzade, Y.; Darabi, F. Surrounded Bonnor Vaidya solution by cosmological fields. \emph{Eur. Phys. J. C} \textbf{2018}, \emph{78}, 1004.
\bibitem{bib:reddy2015impact}     Reddy, K.P.; Govender, M.; Maharaj,S.D. Impact of anisotropic stresses during dissipative gravitational collapse 
 \emph{Gen. Relativ. Gravit.} \textbf{2015}, \emph{47}, 35.
\bibitem{bib:thirukkanesh2012final}     Thirukkanesh, S.; Moopanar, S.; Govender, M. The final outcome of dissipative collapse in the presence of $\Lambda$. \emph{Pramana J. Phys.} \textbf{2012}, \emph{79}, 223--232.
\bibitem{bib:thirukkanesh2013}    Thirukkanesh, S.; Govender, M. The role of the electromagnetic field in dissipative collapse. \emph{Int. J. Mod. Phys. D} \textbf{2013}, \emph{22}, 1350087.
\bibitem{bib:vunk} Wang, A.; Wu, Y. Generalized Vaidya solutions. \emph{Gen Relativ. Gravit.}  \textbf{1999}, \emph{31}, 107.
\bibitem{bib:husain1996exact}    Husain, V. Exact solutions for null fluid collapse. \emph{Phys. Rev. D} \textbf{1996}, \emph{53}, R1759.
\bibitem{bib:Radiationstring1998}   Glass, E.N.; Krisch, J.P. Radiation and string atmosphere for relativistic stars. \emph{Phys. Rev. D} \textbf{1998}, \emph{57}, 5945.
\bibitem{bib:twofluidatm1999}    Glass, E.N.; Krisch, J.P. 
Classical and Quantum Gravity
Two-fluid atmosphere for relativistic stars. \emph{Class. Quant. Grav.} \textbf{1999}, \mbox{\emph{16}, 1175.}
\bibitem{bib:Maombi1}    Mkenyeleye, M.D.; Goswami, R.; Maharaj, S.D. Vaidya and generalized Vaidya solutions by gravitational de-
coupling. \emph{Phys. Rev. D} \textbf{2015}, \emph{92}, 024041.
\bibitem{bib:Maombi2}    Mkenyeleye, M.D.; Goswami, R.; Maharaj, S.D. Thermodynamics of gravity favours Weak Censorship Conjecture.
\emph{Phys. Rev. D} \textbf{2014}, \emph{90}, 064034.
\bibitem{bib:ver1}Vertogradov, V. The eternal naked singularity formation
in the case of gravitational collapse of generalized Vaidya spacetime. \emph{Int. J. Mod. Phys. A} \textbf{2018}, \emph{33}, 1850102. arXiv:2210.16131 [gr-qc]
 \bibitem{bib:ver2} Vertogradov, V. Naked singularity formation in
generalized Vaidya space-time. \emph{Grav. Cosmol.} \textbf{2016}, \emph{22}, 220--223.
\bibitem{bib:ver3} Vertogradov, V. Gravitational collapse of Vaidya
spacetime. \emph{Int. J. Mod. Phys. Conf. Ser.}
\textbf{2016}, \emph{41}, 1660124.
\bibitem{bib:myrev} Saibal Ray, Arijit Panda, Bivash Majumder, Md. Rabiul Islam, \textit{Collapsing scenario for the {\bf k}-essence emergent generalised Vaidya spacetime in the context of massive gravity's rainbow} Chinese Physics C 46(12) DOI: 10.1088/1674-1137/ac8868
\bibitem{bib:r2} Dey, D.; Joshi, P.S. Gravitational collapse of baryonic and
dark matter. \emph{Arab. J. Math.} \textbf{2019}, \emph{8}, 269.
\bibitem{bib:maharaj}   Samson Ojako et al 2020 Class. Quantum Grav. 37 055005 
\bibitem{bib:r4}  Nikolaev, A.V.; Maharaj, S.D. Embedding with Vaidya geometry. \emph{Eur. Phys. J. C} \textbf{2020}, \emph{80}, 648.
\bibitem{bib:r1} Faraoni, V.; Giusti, A.; Fahim, B.H.
 Vaidya geometries and scalar fields with
null gradients.  \emph{Eur. Phys. J. C} \textbf{2021}, \emph{81}, 232.
\bibitem{bib:r3}Brassel, B.P.; Maharaj, S.D.; Goswami, R. Charged radiation collapse in Einstein Gauss Bonnet gravity.  \emph{Eur. Phys. J. C} \textbf{2022}, \mbox{\emph{82}, 359.}
\bibitem{bib:verneg}  V. Vertogradov, Universe 6(9), 155 (2020), arXiv:2209.10976 (gr-qc)
\bibitem{bib:verforce} Vitalii Vertogradov, \textit{Forces in Schwarzschild, Vaidya and generalized Vaidya spacetimes}, 2021 J. Phys.: Conf. Ser. 2081 012036
DOI 10.1088/1742-6596/2081/1/012036
\bibitem{bib:verlinear} V.D. Vertogradov, Non-linearity of Vaidya spacetime and forces in the central naked singularity. Physics of Complex Systems, 2022, vol. 3, no. 2 [arXiv:2203.05270]
\bibitem{bib:vernew} Vitalii Vertogradov, Maxim Misyura "Vaidya and Generalized Vaidya
Solutions by Gravitational Decoupling"Universe 2022, 8(11), 567;
doi:10.3390/universe8110567 arXiv:2209.07441 [gr-qc]
\bibitem{bib:quasi} Badri Krishnan, Quasi-local black hole horizons. [arXiv:1303.4635v1 [gr-qc]]
\bibitem{bib:newin} Ayon Tarafdar, Srijit Bhattacharjee, Slowly evolving horizons in Einstein gravity and beyond. [arXiv:2210.15246 [gr-qc]]
 \bibitem{bib:hawkingradation}S. W. Hawking (1975), Particle Creation by Black Holes, Comm. Math. Phys. 43, 199.
\bibitem{bib:hok} Hawking, S.W.; Ellis, G.F.R. \emph{The Large Scale
Structure of Space-Time}; Cambridge University Press: Cambridge, UK, 1973.
\bibitem{bib:pois} Poisson, E. \emph{A Relativist's Toolkit: The Mathematics of
Black-Hole Mechanics}; Cambridge University Press: Cambridge, UK, 2007.
			\bibitem{bib:zel} Ya.B. Zel'dovich, I.D. Novikov Theory of gravitation and the evolution of stars, Moskva: Nauka, 1971, 484 p.
\bibitem{bib:etnak} Vitalii Vertogradov. The structure of the generalized Vaidya spacetime containing the eternal naked singularity. International Journal of Modern Physics A Vol. 37, No. 28n29, 2250185 (2022) [arXiv:2209.10953.]
\bibitem{bib:nelspat} Nielsen, A.B. The Spatial relation between the event horizon and trapping horizon. Class. Quant. Gravity 2010, 27, 245016, doi:10.1088/0264-9381/27/24/245016.
\bibitem{bib:frommahhomvay}J. Lewandowski, Class. Quant. Grav, 7, L135 , (1990).
`\bibitem{bib:13} G. Fodor, K. Nakamura, Y. Oshiro and A. Tomimatsu, Surface gravity in dynamical spherically-symmetric spacetimes, Phys. Rev. D 54, 3882 (1996) [arXiv:gr-qc/9603034].
\bibitem{bib:kodama} H. Kodama, Conserved Energy Flux For The Spherically Symmetric System And The Back Reaction Problem In The Black Hole Evaporation, Prog. Theor. Phys. 63, 1217 (1980).
\bibitem{bib:14} S. A. Hayward, Unified first law of black-hole dynamics and relativistic thermodynamics, Class. Quant. Grav. 15 (1998) 3147 [arXiv:gr-qc/9710089].
\bibitem{bib:11} A. B. Nielsen and M. Visser, Production and decay of evolving horizons, Class. Quant. Grav. 23 (2006) 4637 [arXiv:gr-qc/0510083].
\bibitem{bib:31ber}R. W. Lindquist, R. A. Schwartz, C. W. Misner, Physical Review 137, 1364 (1965).
\bibitem{bib:berezin} V. A. Berezin, V. I. Dokuchaev, Yu. N. Eroshenko \textit{Vaidya Spacetime in the Diagonal Coordinates}, JETP 124, 446 (2017)
\bibitem{bib:verdiag} Vitalii Vertogradov \textit{The diagonalization of generalized Vaidya spacetime }, International Journal of Modern Physics A 2020
\bibitem{bib:wisser}Visser, M. Dirty black holes: Thermodynamics and
horizon structure. \emph{Phys. Rev. D} \textbf{1992}, \emph{46}, 2445--2451.
\bibitem{bib:124}    R. W. Lindquist, R. A. Schwartz, C. W. Misner, Vaidyas Radiating Schwarzschild Metric. Phys. Rev., 137(5B), B1364 (1965).
\bibitem{bib:charged1} W. B. Bonnor and P. C. Vaidya, "`Spherically symmetric radiation of charge in Einstein-Maxwell theory"' Gen. Rel. Grav. 1, 127 (1970).
\bibitem{bib:charged2}Chatterjee, S., Ganguli, S. and Virmani, A. Charged Vaidya solution satisfies weak energy condition. Gen Relativ Gravit 48, 91 (2016), arXiv:1512.02422 (gr-qc)
\bibitem{bib:force} Jiri Kovar, Zdenek Stuchlik. Forces in Kerr spacetimes with a repulsive cosmological constant.Int.J.Mod.Phys.A21:4869-4897,2006   [arXiv:gr-qc/0611152 ]
\bibitem{bib:parth} Parth Bambhaniya, Dipanjan Dey, Ashok B. Joshi, Pankaj S. Joshi, Divyesh N. Solanki, Aadarsh Mehta. Shadows and negative precession in non-Kerr spacetime. Phys. Rev. D 103, 084005 (2021) [arXiv:2101.03865 [gr-qc]]
\end{thebibliography}
\end{document}